\documentclass[twocolumn]{aastex62}

\received{January 1, 2018}
\revised{January 7, 2018}
\accepted{\today}
\submitjournal{ApJ}

\shorttitle{The Hard X-ray CSPN of the Eskimo Nebula}
\shortauthors{Guerrero et al.}

\begin{document}


\title{
Variable Hard X-ray Emission From the Central Star of the Eskimo Nebula}


\correspondingauthor{Mart\'\i n A. Guerrero}
\email{mar@iaa.es}

\author[0000-0002-7759-106X]{Mart\'\i n A. Guerrero}
\affil{Instituto de Astrof\'isica de Andaluc\'ia, IAA-CSIC, 
Glorieta de la Astronom\'\i a s/n, 18008 Granada, Spain}

\author[0000-0002-5406-0813]{Jes\'us A.\ Toal\'a}
\affiliation{Instituto de Radioastronom\'\i ıa y Astrof\'\i sica (IRyA), 
UNAM Campus Morelia, Apartado postal 3-72, 58090 Morelia, Michoacan, Mexico}

\author[0000-0003-3667-574X]{You-Hua Chu}
\affiliation{Institute of Astronomy and Astrophysics, Academia Sinica 
(ASIAA), Taipei 10617, Taiwan, Republic of China}

\begin{abstract}

The central star of NGC\,2392 shows the hardest X-ray 
emission among central stars of planetary nebulae (CSPNe).  
The recent discovery of a spectroscopic companion with an orbital period of 
1.9 days could provide an explanation for its hard X-ray emission, as well 
as for the collimation of its fast outflow.  
Here we analyse the available \emph{Chandra} and \emph{XMM-Newton} 
X-ray observations to determine accurately the spectral and temporal 
variation properties of the CSPN of NGC\,2392.  
The X-ray emission can be described by an absorbed thermal plasma 
model with temperature 26$^{+8}_{-5}$ MK and X-ray luminosity 
(8.7$\pm$1.0)$\times$10$^{30}$ erg~s$^{-1}$.  
No long-term variability is detected in the X-ray emission level, 
but the \emph{Chandra} light curve is suggestive of short-term 
variations with a period $\sim$0.26 days.  
The possible origins of this X-ray emission are discussed.  
X-ray emission from the coronal activity of a 
companion or shocks in the stellar wind can be 
ruled out.  
Accretion of material from an unseen main-sequence companion onto the CSPN 
or from the CSPN wind onto a white dwarf companion are the most plausible 
origins for its hard X-ray emission, although the mismatch between the 
rotational period of the CSPN and the modulation time-scale of the X-ray 
emission seems to preclude the former possibility.  

\end{abstract}


\keywords{ISM: planetary nebulae: general -- 
          ISM: planetary nebulae: individual: NGC\,2392 -- 
          stars: winds, outflows --
          (stars:) white dwarfs --
          X-rays: stars -- 
          X-rays: individual (NGC\,2392)
          }

\section{Introduction}

The planetary nebula (PN) NGC\,2392, the Eskimo Nebula, shows a 
double shell morphology with an inner elliptical shell displaying an 
ensemble of filaments and a round outer shell where a series of wisps 
resembling the fur on a hood can be seen \citep{O'Dell_etal02}.  
This morphology is somewhat peculiar among PNe, but it is 
not completely unseen \citep[e.g., NGC\,7662,][]{GJC2004}.  
Actually, its three-dimensional structure might be similar to 
those of NGC\,6543 and NGC\,7009, but at a different viewing 
angle \citep{GDetal_2012}.

Everything else in NGC\,2392 is atypical.  
The expansion velocity of its inner shell, 
$\sim$120 km~s$^{-1}$, is spectacularly large 
among elliptical shells of PNe 
\citep{RAT83,OB85,OWC90,GDetal_2012}.  
Its fast $\sim$180 km~s$^{-1}$ collimated outflow, the first ever 
detected among PNe \citep{GBS85}, can be traced down to its central 
star (CSPN), a situation uncommon among collimated outflows of 
PNe.  
Its central star exhibits the hardest X-ray emission among CSPNe 
\citep{Retal_2013,Metal2015}.  
The nebular excitation implied by the bright He~{\sc ii} emission relative 
to H$\beta$ \citep{Detal_2012}, the He~{\sc ii} Zanstra temperature 
of $\sim$70,000~K and even higher energy-balance temperature \citep{MKH92}
are too high for the CSPN's effective temperature of 40,000-45,000 K  
\citep{PHM04,HB11}.
Furthermore, the difference between the electron temperatures derived
from [O~{\sc iii}] and [N~{\sc ii}] lines is much larger than usual \citep{Barker91}.

All these unusual properties are suggestive of the presence of an 
additional source of He ionizing photons and mechanical momentum 
within the inner nebular shell of NGC\,2392.  
The fast outflow and the hard X-ray emission from the CSPN, in 
conjunction with the diffuse soft X-ray emission from hot plasma 
confined within the inner shell \citep{Getal05,Retal_2013},
have been suggested to provide these additional sources of 
momentum and ionizing photons \citep{Ercolano2009}.  
Alternatively, a hot ($\simeq$250,000 K), high-gravity white dwarf (WD) 
binary companion has been proposed for NGC\,2392 \citep{Detal_2012}, 
and the recent detection of a spectroscopic binary with an orbital 
period of 1.9 days support this suggestion \citep{Metal_2019}.
The presence of a binary system at the CSPN of NGC\,2392 may 
finally explain the origin of its hard X-ray emission from
the accretion of the wind of the CSPN onto the WD companion
\citep{Metal_2019}, and can also be linked to the currently
active collimation and launch of the fast bipolar outflow,
which might reveal itself through its fast polar wind 
\citep{PU2014}.

To investigate the link between accretion and the hard X-ray
emission from the CSPN of NGC\,2392, we present in this paper
a spectral and temporal variation analysis of pointed 
\emph{Chandra} and \emph{XMM-Newton} observations.  
These observations are supplemented with multi-wavelength 
IR, optical, and UV observations and/or archival data to 
build a spectral energy distribution (SED) of the CSPN of 
NGC\,2392.

\section{Observations and Data}


\subsection{X-ray Observations}

The \emph{XMM-Newton} Observatory observed NGC\,2392 in Revolution 790
on 2004 April 2 (Obs.ID.: 0200240301; PI: Y.-H.\,Chu) using the
European Photon Imaging Cameras (EPIC). 
The two EPIC-MOS cameras were operated in the Full Frame mode, 
while the EPIC-pn camera was operated in the Extended Full 
Frame mode. 
The total observing time was 17.7~ks for both MOS 
and 14.3~ks for pn.  
All observations were obtained with the medium optical blocking filter.
The {\it XMM-Newton} pipeline products were processed using 
the {\it XMM-Newton} Science Analysis Software (SAS) and the 
calibration files from the Calibration Access Layer available 
on 2019 July 1. 
The event files were screened to eliminate events due to charged particles 
or associated with periods of high background.  
For the EPIC-MOS observations, only events with CCD patterns 0–12 
were selected, whereas for the EPIC-pn observation only single pixel
events were selected.  
Time intervals with high background count rates in 
the background-dominated 10--12 keV energy range
(i.e., $\geq$0.2 counts~s$^{-1}$ for MOS and 
$\geq$0.45~counts~s$^{-1}$ for pn) were discarded. 
The resulting exposure times are 17.3~ks, 17.5 ks, and 8.3 ks for 
MOS1, MOS2, and pn, respectively.

The \emph{Chandra} X-ray Observatory observed NGC\,2392 
for 57.4~ks on 2007 September 13 (Obs.ID: 7421; PI: M.A.\ 
Guerrero).
The array for spectroscopy of the Advanced CCD Imaging Spectrometer
(ACIS-S) was used and the PN was imaged on the back-illuminated CCD S3
using the VFAINT mode. 
The data were processed and analysed using the \emph{Chandra}
Interactive Analysis of Observations (CIAO) software package 
\citep[version 4.11;][]{Fruscione2006}.
The observations were not affected by high-background events, but periods of 
time with background count rates above 0.44~counts~s$^ {-1}$ or anomalously 
low at the beginning or end of the observation were excised to allow a fair 
investigation of the temporal variation of NGC\,2392.  
The final exposure time after this procedure was 50.4~ks.

\subsection{Complementary Observations and Archival Data}

\subsubsection{IR Observations}

Near-IR $JHK$ images and long-slit spectra 
were obtained 
with the Near Infrared Camera Spectrometer (NICS) on the 3.5~m Telescopio 
Nazionale Galileo (TNG) at the Observatorio de El Roque de los Muchachos 
(ORM) on the island of La Palma, Spain.  
The detector was a HgCdTe Hawaii 1024$\times$1024 array.  
The large field camera was used for the imaging observations, yielding 
a pixel size of 0\farcs25 and a field-of-view of 4\farcm2$\times$4\farcm2.  
The medium-resolution $J$, $H$, and $KB$ prisms were used in conjunction
with a 0\farcs75 slitwidth for the spectroscopic observations, resulting 
in a spectral resolution of $\sim$1,200.


We have also used archival mid-IR \emph{Spitzer Space Telescope} 
Infrared Array Camera \citep[IRAC;][]{Fazio_etal04} and Infrared 
Spectrograph \citep[IRS;][]{Hetal_2004} observations of NGC\,2392.  
The \emph{Spitzer} IRAC observations (Prog.\ ID 30285, PI G.\ Fazio) 
consisted of images in the 3.6, 4.5, 5.8, and 8.0 $\mu$m bands.  
The \emph{Spitzer} IRS observations (Prog.\ ID 30482, PI J.R.\ Houck) 
were obtained in Spectral Mapping mode to provide multiple spectra 
from a 2-D region encompassing the inner shell and the surrounding 
outer shell of NGC\,2392.  
A spectrum of the CSPN was extracted in the wavelength range 
from 5.3 to 14.7 $\mu$m.

\subsubsection{Optical and UV Observations}

$U$, $B$, and $R$ magnitudes of the CSPN of NGC\,2392 were adopted from 
SIMBAD, and $V$ and $I$ magnitudes from \citet{Ciardullo_etal99}.  
The \emph{Far-Ultraviolet Spectroscopic Explorer} (\emph{FUSE}) dataset
B0320601 and the \emph{International Ultraviolet Explorer} (\emph{IUE})
datasets SWP05230LL and LWR04209LL have been used, providing sensitive
observations in the spectral ranges 905--1185 \AA\ and 1150--3350 \AA,
respectively.

\begin{figure*}[!t]
\centerline{\includegraphics[width=1.80\columnwidth,angle=0]{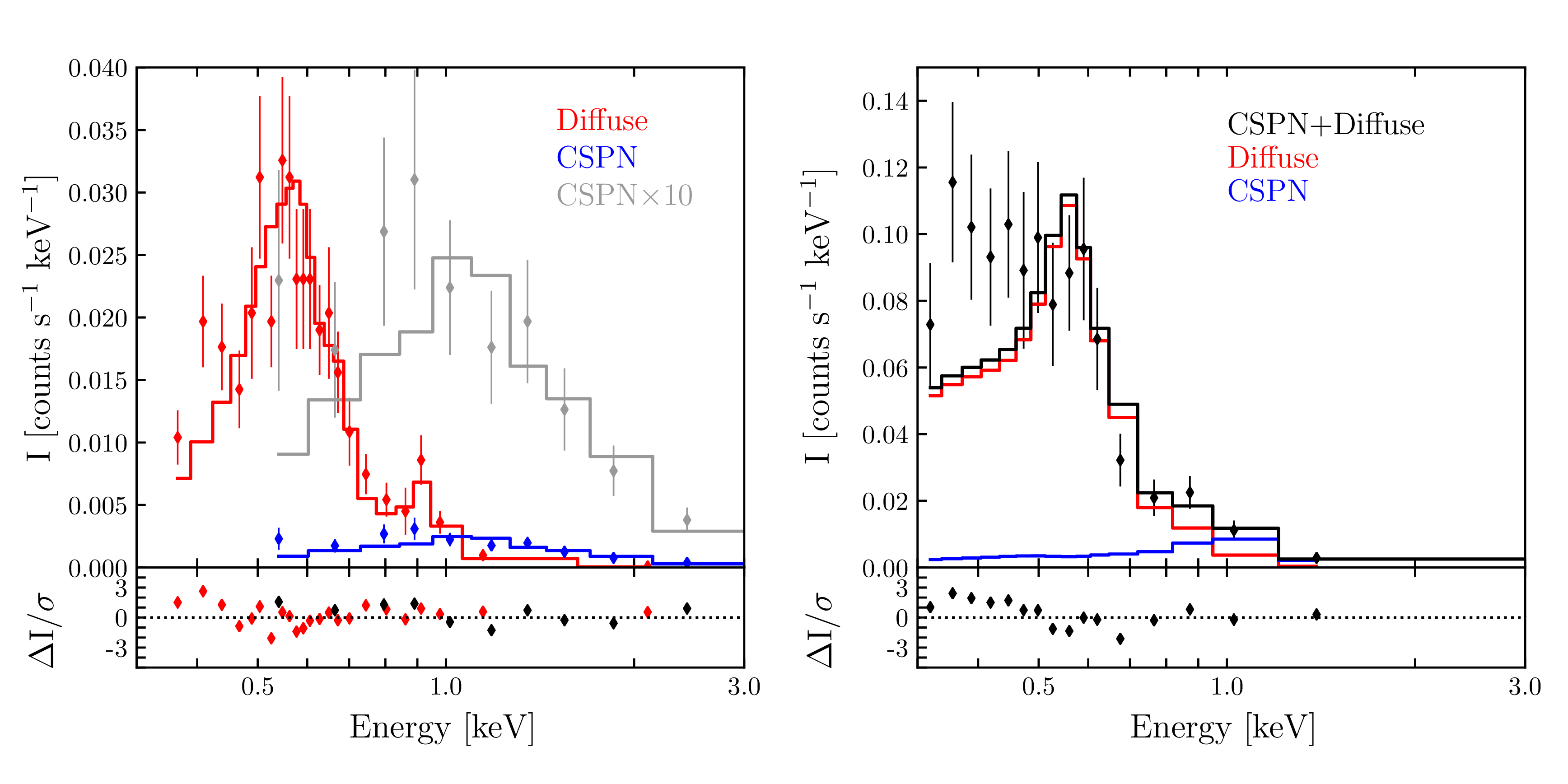}}
\caption{
\emph{Chandra} ACIS-S S3 (left) and \emph{XMM-Newton} EPIC-pn (right) 
background-subtracted spectra (dots) of the CSPN of NGC\,2392 and its 
diffuse X-ray emission.  
The X-ray spectra are overplotted with the best-fit models (solid histograms).  
The ACIS-S3 spectrum of the CSPN of NGC\,2392 and its best-fit model 
are shown at two different scales to display more clearly its spectral 
shape and the relative intensities of the CSPN and diffuse emissions.  
Residuals of the fits are plotted in the lower panels.  
Both the \emph{Chandra} ACIS-S3 and \emph{XMM-Newton} EPIC-pn 
spectra of the diffuse emission show residuals in the 0.3--0.45 
KeV energy range, which might be attributed to emission from the 
C~{\sc vi} 0.37 keV and N~{\sc vi} 0.43 keV lines, thus implying 
higher carbon and nitrogen abundances than those of the stellar 
wind considered here.   
}
\label{xspec}
\end{figure*}

\section{Hard X-rays of NGC\,2392 CSPN}

The \emph{Chandra} observations of NGC\,2392 detect a point-like 
X-ray source at its CSPN, as well as diffuse X-ray emission within
the inner nebular shell \citep{Retal_2013,Metal2015}.  
The source at the CSPN of NGC\,2392 has a full width at half maximum (FWHM)
of 0\farcs44$\pm$0\farcs06 comparable to that of the point spread function 
(PSF) of \emph{Chandra} at the ACIS-S aimpoint.  
Thus, its point-source nature is confirmed.  
The positional coincidence between the X-ray point-source and the 
CSPN of NGC\,2392 merits careful investigation.  
The \emph{Chandra} ACIS-S S3 field of view includes 5 point sources
with stellar counterparts whose coordinates are available in the USNO 
2.0 Catalog.  
Using the coordinates of these 5 stars, the X-ray and optical positions 
were registered within 0\farcs4.  
The positions of the X-ray point-source detected by \emph{Chandra} 
near the center of NGC\,2392 and the CSPN are coincident within
0\farcs2.  
Clearly, the X-ray point source is not associated with the visual companion 
at 2\farcs65 S-SW of the CSPN \citep{Ciardullo_etal99}.  It is also very 
unlikely that a foreground star or a background extragalactic source is 
located exactly along the line of sight 
toward the CSPN and is responsible for the X-ray point source.
It can thus be confidently concluded that the X-ray emission from this
point source originates from the CSPN of NGC\,2392 or from a source in 
its close vicinity.

To study the spectral properties and possible variations of the X-ray emission
from the CSPN of NGC\,2392, we have defined a source region of radius 1\farcs5
centered at its location and a surrounding 12\farcs8$\times$16\farcs0 
elliptical background region representative of the diffuse emission.  
The point-source at the CSPN of NGC\,2392 has a background-subtracted 
count rate of 3.3$\pm$0.3 ACIS-S S3 counts~ks$^{-1}$ for a total of 
185$\pm$15 counts.  
Its X-ray spectrum, shown in black in the left panel of Figure~\ref{xspec}, 
peaks at 0.8--1.0 keV and shows a hard-energy tail that declines steadily 
to 3 keV, with some residual emission at energies as high as to 4.0 keV.  
This X-ray spectrum is much harder than that of the diffuse
emission \citep[shown in red in the left panel of
Fig.~\ref{xspec}-{\it left},][]{Getal05}, which peaks at
0.5--0.6 keV and diminishes at energies above 1 keV.  
Assuming a uniform surface brightness for the diffuse emission, 
the total diffuse emission within the aperture of the CSPN is
$\leq$12\% as high as the background-subtracted point source
emission.  
We have tested different background regions representative of the diffuse
emission and found that the X-ray spectrum of the CSPN is unaffected at
energies $\geq$1.0 keV, but the contamination of diffuse emission may
produce variations up to $\simeq$10\% at energies between 0.6 and 0.9 KeV,
and up to $\simeq$40\% at energies below 0.5 keV.  
These results are similar to those presented by \cite{Metal2015}.

The likely presence of emission lines between 0.8 and 1.0 keV in
the spectrum of the CSPN suggests that it can be interpreted as
emission produced by an optically thin plasma.  
Despite the small number of counts, we have used XSPEC \citep{Arnaud1996} 
to obtain a rough fit of this spectrum using an absorbed APEC thin plasma 
emission model\footnote{
We note here that the spectrum of the CSPN of NGC\,2392 can also be 
fitted by a highly absorbed power law, but this seems inconsistent 
with the low hydrogen column density derived from optical and UV 
observations.
}
assumed to have the subsolar $\lesssim$0.25 $Z_\odot$ stellar
abundances derived by \citet{HB11} and an absorption column
density $N_{\rm H}$ of 9$\times$10$^{20}$ cm$^{-2}$ corresponding
to its optical extinction, $c_{{\rm H}\beta}$=0.225 \citep{PB-SR08}.  
The best-fit model ($\chi^2$/DoF=10.6/9=1.2) with a temperature 
$kT$=2.2$^{+0.7}_{-0.4}$ keV is overplotted on the background-subtracted 
X-ray spectrum in Figure~\ref{xspec}-{\it left}.  
This temperature is lower than but consistent with that of 
3.1$^{+1.7}_{-0.7}$ KeV derived by \citet{Metal2015} 
assuming solar abundances.  
If a solar abundance were assumed instead, the temperature of the
best fit would rise up to 5.4$\pm$2.3 keV, but the quality of the
fit would worsen ($\chi^2$/DoF$\simeq$1.6). 
The absorbed X-ray flux in the 0.3--3.0 keV band is 1.3$\times$10$^{-14}$ 
erg~cm$^{-2}$~s$^{-1}$.  
The \emph{Hipparcos} distance to NGC\,2392 of 1150 pc 
\citep{Petal_1997} is consistent with the expansion 
distance of 1300 pc derived by \citet{GDetal_2015}, 
but the most recent \emph{Gaia} determination implies 
a larger distance of 2000$\pm$200 pc \citep{KB2018}.  
Using the latter, we derive an intrinsic luminosity in the 0.3--3.0 
energy band of (8.7$\pm$1.0)$\times$10$^{30}$ erg~s$^{-1}$.  

The diffuse X-ray emission of NGC\,2392 is spatially unresolved from that 
of its CSPN in the \emph{XMM-Newton} EPIC observations \citep{Getal05}, 
but its spectral shape is well described by the \emph{Chandra} ACIS-S S3 
spectrum.  
This can be fitted ($\chi^2$/DoF=24.8/21=1.2) using an optically 
thin plasma APEC emission model with nebular abundances and 
absorption column density \citep{PB-SR08} for a temperature 
$kT$=0.16$^{+0.015}_{-0.009}$ keV (Fig.~\ref{xspec}-{\it left}).  
The X-ray luminosity of this diffuse emission is 
(7.2$\pm$0.6)$\times$10$^{31}$ erg~s$^{-1}$.  
This model for diffuse emission is adopted for the \emph{XMM-Newton} EPIC-pn 
spectrum (red color in Figure~\ref{xspec}-{\it right}) to obtain a net hard 
X-ray excess that can be attributed to the emission from the CSPN.  
This component has been fit with a similar APEC model 
to that used for the \emph{Chandra} ACIS-S S3 spectrum 
of the CSPN of NGC\,2392.  
The best-fit model ($\chi^2$/DoF=33.4/24=1.4) with a temperature 
$kT$=1.2$^{+3.3}_{-0.3}$ keV is overplotted on the background-subtracted 
X-ray spectrum in Figure~\ref{xspec}-{\it right} (blue histogram).  
The absorbed X-ray flux in the 0.3--3.0 keV band is 
8.5$\times$10$^{-15}$ erg~cm$^{-2}$~s$^{-1}$ and the 
intrinsic luminosity is (7$\pm$3)$\times$10$^{30}$ erg~s$^{-1}$, 
which is consistent with that derived from \emph{Chandra}.


\begin{figure}[!t]
\centerline{\includegraphics[bb=18 220 582 712,width=0.975\columnwidth,angle=0]{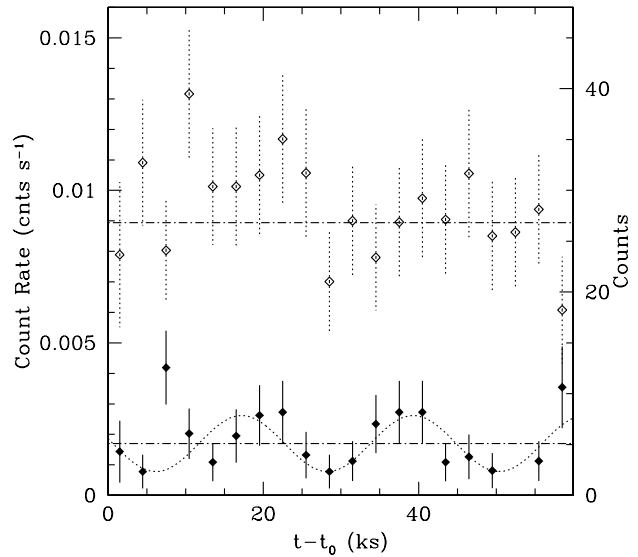}}
\caption{
\emph{Chandra} ACIS-S S3 light curve of the CSPN of NGC\,2392 
(solid diamonds) in the energy range 1.0--3.0 keV.  
The light curve of the diffuse nebular emission in the 0.3--2.0 keV 
band, which is expected to be constant, is also shown for comparison 
(open diamonds).  
The dash-dotted horizontal lines in the light curves mark the averaged count 
rates.  
The dotted line corresponds to the best-fit sinusoidal model.  
}
\label{xlc}
\end{figure}

The ACIS-S S3 light curve of the CSPN of NGC\,2392 in the X-ray 
energy range 1.0--3.0 keV free from diffuse emission is shown 
in Figure~\ref{xlc}.  
This X-ray light curve shows some oscillations, very different from the 
behavior of the light curve of the nebular emission, which is expected 
to be constant and indeed shows little variations around the mean value 
all through the duration of the observation.  
To test for variability, we performed a Kolmogorov-Smirnov test 
to assess whether these light curves can be ascribed to a Poisson 
distribution of the photon arrival times.  
There is a 97\% probability that the time arrival of the diffuse X-ray 
emission follows a Poisson distribution, but only 50\% for the CSPN.  
This same test was applied to three point-sources 
(2CXO\,J072908.2+205609, 
2CXO\,J072912.1+210113, and 
2CXO\,J072918.8+205551) in the field of view of ACIS-S S3 
with similar count rates to those of the CSPN of NGC\,2392, 
resulting always in probabilities for Poisson distributions 
of the photon arrival times $\gtrsim$85\%.  
Therefore, the X-ray variability of the hard X-ray source at the 
CSPN of NGC\,2392, although not firm, is suggested by the shape 
of its light curve.
A $\chi^2$ fit of a sinusoidal curve to this light curve
(Fig.~\ref{xlc}) is indicative of a period $\simeq$22 ks 
($\simeq$0.26 days).   
This fit has a reduced $\chi^2$ value of 1.1.  
For comparison, a similar fit to a constant value has a reduced
$\chi^2$  value of 1.5.
This period is consistent with the period $\simeq$0.253 days derived from
the periodogram obtained using the Lomb-Scargle method,
although we reckon that the false alarm probability of 15\% which is derived
constitutes only a weak indication of periodicity. 

\section{Discussion}

Three possible origins exist for the X-ray emission from the CSPN of 
NGC\,2392 \citep[see, for instance,][]{Metal2015}.  
They are discussed below.

\subsection{Coronal Activity from a Companion}

\begin{figure}[!t]
\includegraphics[bb=64 185 540 700,width=0.95\columnwidth,angle=0]{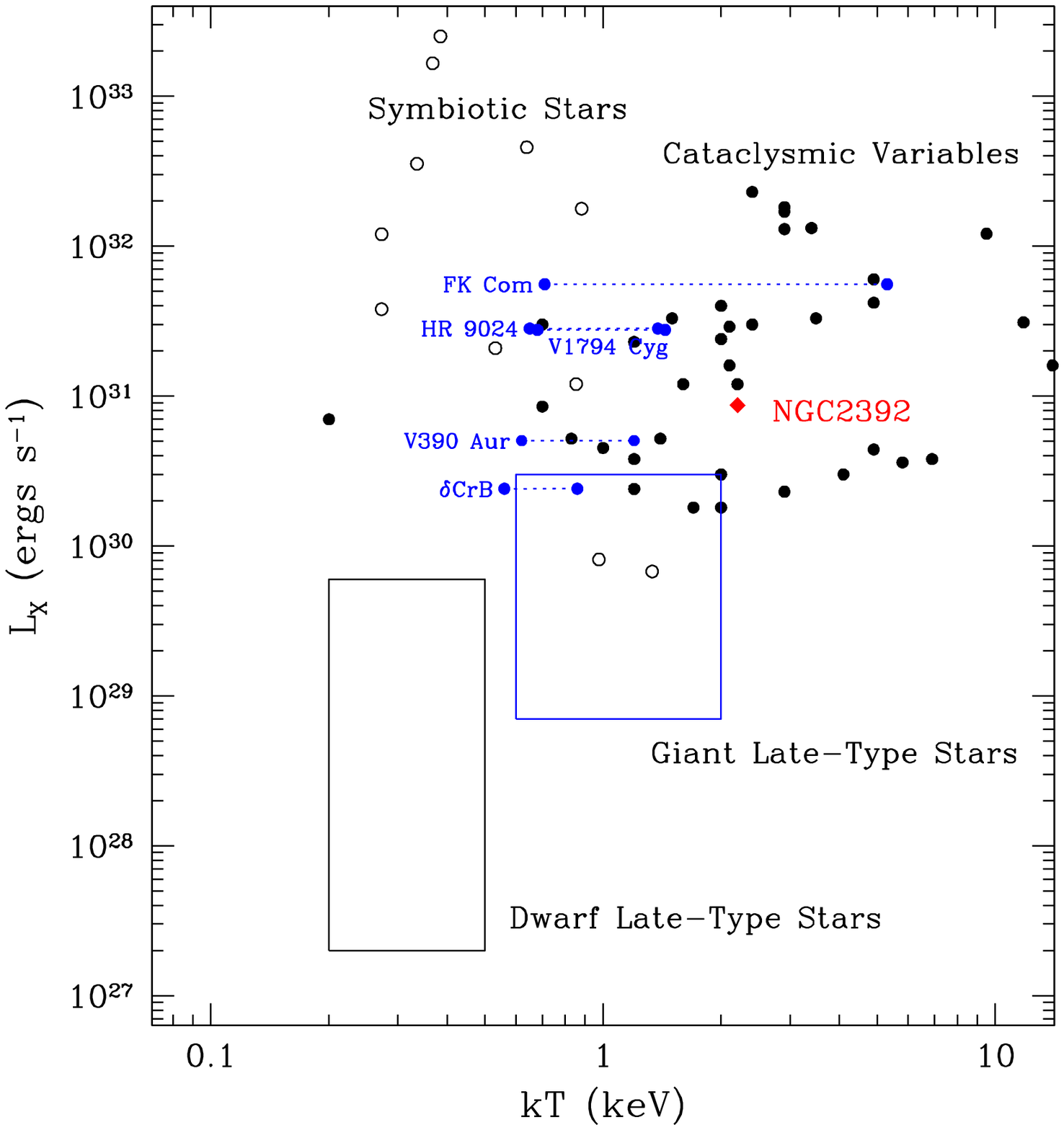}
\caption{
Comparison of the location on the $L_{\rm X}$ vs.\ $kT$ plane of the CSPN 
of NGC\,2392 with these of active stars, cataclysmic variables, and symbiotic 
stars.   
The averaged location of dwarf and giant late-type stars 
\citep{Maggio_etal90,HSV98a,HSV98b} are marked by black 
and blue boxes, respectively.  
Individual rapidly rotating G giant stars with accurate determinations of 
their X-ray luminosities using two-temperature thermal plasma emission
models are marked by solid blue dots at the temperature of each thermal
component connected by dotted lines
\citep{Gondoin_etal2002,Gondoin2003a,Gondoin2003b,Gondoin2004,Gondoin2005}. 
Individual cataclysmic variables \citep{EHP91} and symbiotic 
stars \citep{MWJ97,SS09,ELS10} are marked by filled and open 
dots, respectively.  
}
\label{fig_lxkt}
\end{figure}

The X-ray emission from the CSPN of NGC\,2392 may originate from the coronal
activity of an unresolved late-type companion \citep{Getal01,Metal_2010} 
whose rotation (and hence its magnetic activity) may have been boosted by the 
transfer of orbital angular momentum \citep{SK02}.
For late-type dwarf stars of spectral types K and M, the value of 
$\log(L_{\rm X}/L_{\rm bol})$ is typically --5.21, although it can 
increase up to --3.56 for saturated activity \citep{FSG95}.  
Main-sequence stars of earlier spectral types, from A0 to K5, show 
similar typical and saturated values for $\log(L_{\rm X}/L_{\rm bol})$ 
\citep{HSV98a,Gudel04}.  
As for giant and sub-giant stars, the relationship between spectral 
type and coronal activity is more complex than for their main sequence 
counterparts \citep[see the review by][]{Gudel04}.  
Giant and sub-giant stars of spectral types later than 
K1 and earlier than G5 are usually weak X-ray emitters 
\citep[e.g.,][]{Ayres_etal81,Haisch_etal90,Maggio_etal90}, 
with the brightest X-ray sources among G and K stars at 
$L_{\rm X}\le$10$^{31}$ erg~s$^{-1}$ and 
$\log(L_{\rm X}/L_{\rm bol})<-3$ \citep{Gondoin00}, 
although some rapidly rotating G giants with permanent flaring activity 
can reach X-ray luminosities up to several times 10$^{31}$ erg~s$^{-1}$ 
\citep[][and references therein]{Gondoin05}.  
A comparison of the plasma temperatures and levels of X-ray
emission from the CSPN of NGC\,2392 and late type stars is
presented in Figure~\ref{fig_lxkt}.
The plasma temperature of the CSPN of NGC\,2392 is marginally consistent
with those of the coronae of giant late type stars \citep{HSSV99}, but
its X-ray luminosity is significantly higher.  
We note that rapidly rotating G giants present similar levels of X-ray 
emission and, for the case of FK\,Com, even higher plasma temperatures 
for the component associated with flares \citep{Gondoin_etal2002},  
but the modulation observed in the X-ray light curve of NGC\,2392 is
too fast to be associated with the rotational period of a late-type
companion.

\subsection{Stellar Wind}

A fast stellar wind can generate shocks capable of producing 
X-ray emission.   
The shock-in-wind origin for the hard X-ray emission of the CSPN of NGC\,2392
is strengthened by the notable variability of its stellar wind \citep{PU2014},
the intriguing presence of the O~{\sc vi} UV resonance doublet 
\citep{HB11,GM2013}, and an X-ray to bolometric luminosity 
$\log(L_{\rm X}/L_{\rm bol})=-6.4$ 
\citep[for an $L_{\rm bol}$ of 6025 $L_\odot$,][]{PHM04,HB11} 
consistent with the value of $-6.6$ observed in shocks in
winds of O stars \citep{Nebot2018}.  
However, the efficiency of conversion of the wind mechanical energy into 
X-rays for OB stars, in the narrow range $-3.1<\log(L_{\rm X})/L_{\rm w})<-2.5$ 
\citep{Sana_etal06}, is clearly violated in NGC\,2392, with a 
$\log(L_{\rm X})/L_{\rm w})=-1.8$, for a wind luminosity of
5.4$\times$10$^{32}$ erg~s$^{-1}$ adopting a terminal wind
velocity of 300 km~s$^{-1}$ and a mass-loss rate of  
1.9$\times$10$^{-8}$ $M_\odot$ yr$^{-1}$ \citep{HB11}.
More importantly, its stellar wind terminal velocity simply cannot
produce hot plasma at a temperature $kT\sim$2.2 keV.

In an OB or WR binary system, colliding stellar winds can produce 
shock-heated plasma at 10$^7$ K ($kT\sim$1 keV).  
The X-ray emission from these systems is characterized by large hydrogen 
absorption column densities and variability on time-scales of orbital period 
caused by the varying amount of un-shocked attenuating material along the 
line of sight.  
However, none of these fit the observed X-ray emission from the
CSPN of NGC\,2392.

On a smaller scale, colliding winds are also typical sources of 
hard X-ray emission in symbiotic stars (SS), interacting binary 
systems formed by a red giant and a hot (degenerate) companion.  
The shape of the X-ray spectra of the so-called class $\beta$ SS is 
particularly similar to that of the CSPN of NGC\,2392; however, their 
X-ray luminosities are higher because the large mass-loss rate of 
the red giant star in these systems \citep{MWJ97}.  
Only the ``low X-ray luminosity SS'' such as EG\,And, SS73\,17, and MWC\,560 
show similar emission levels to those of the CSPN of NGC\,2392, but the 
temperature of the X-ray emitting plasma in these systems is notably lower. 

\begin{figure}[!t]
\includegraphics[bb=35 240 550 570,width=1.0\columnwidth,angle=0]{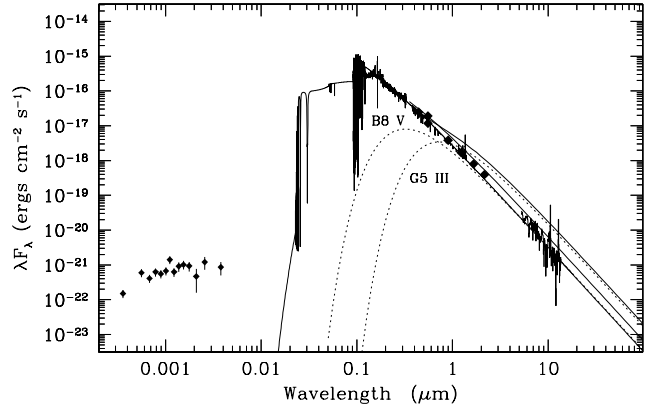}
\caption{
X-ray to IR spectral energy distribution (SED) of the CSPN of NGC\,2392.
The data points and/or spectra have been corrected for extinction
to derive the intrinsic SEDs using \citet{CCM89}'s extinction law
in the IR, optical, and UV.
The emission expected from the star is represented by a NLTE model computed 
using WMAP, the T\"ubingen NLTE Model-Atmosphere Package Web interface, at
http://astro.uni-tuebingen.de/$\sim$rauch \citep{Rauch97} at a temperature
of 45,000~K, a stellar radius of 0.7 $R_\odot$, and a distance of 2,000~pc
assuming a surface He:H ratio of 0.1 by number (thin solid line).  
The contribution from a B8~V or a G5~III companion 
(dotted lines) would be detectable in the near- and
mid-IR SED of the CSPN of NGC\,2392, as shown by
the addition of the two spectral components as a
thick solid line.  
}
\label{sed}
\end{figure}

\subsection{Accretion}

Accretion onto the CSPN or an accretion disk from an unseen companion can
also result in hard X-ray emission \citep{SL94,MM98,R-RL99}, e.g., as in
quiescent novae and cataclysmic variables (CVs), where strong UV and X-ray
emission is typically detected.  
Such companion would be a main-sequence star of spectral type F0~V
or later, as a star with earlier spectral type, such as the B8~V star,
whose emission is represented in Figure~\ref{sed} assuming a black-body
model at their temperature and luminosity, would be detectable in the
near- and mid-IR SED of the CSPN of NGC\,2392.
As shown in this figure, a giant star would also be detected at long
wavelengths.  
Quiescent novae have higher X-ray luminosities, typically
$\gtrsim$10$^{32}$ erg~s$^{-1}$, and X-ray spectra harder
than that of the CSPN of NGC\,2392 \citep{OCO2001}, but
CVs have similar X-ray luminosities and plasma temperatures
as shown in Figure~\ref{fig_lxkt}.  
Quiescent novae and CVs, however, are known to exhibit 
large variability in the optical and X-rays, which do
not match those of the CSPN of NGC\,2392.  
Therefore, accretion processes at rates larger than 
10$^{-12}$ $M_\odot$~yr$^{-1}$, typical of quiescent 
novae or CVs, can be ruled out as the origin of the 
hard X-ray emission in the CSPN of NGC\,2392.  
The modulation of $\sim$0.26 days in the X-ray light curve of NGC\,2392 
is twice its rotational period of 0.123 days \citep{PU2014}, suggesting 
the X-ray emission does not arise from a hot spot on the surface of the 
CSPN.

The recent discovery of a binary companion of the CSPN of NGC\,2392 
provides the tantalizing opportunity to explain the origin of its 
hard X-ray emission from the accretion of the stellar wind of the 
CSPN of NGC\,2392 onto a relatively massive ($M\gtrsim0.6~M_\odot$) 
WD companion or an accretion disk around it \citep{Metal_2019}.  
Whereas an accretion disk around this putative WD companion would not 
produce X-ray emission, for its maximum temperature $\leq$100,000 K 
\citep{Getal01}, accretion directly onto the WD can raise temperatures 
well up to $kT\sim$7.6 keV, well above the observed X-ray temperature 
of 2.2 keV.  
The observed X-ray luminosity demands that only $\sim$1\% of the 
stellar wind from the CSPN of NGC\,2392 \citep[$\dot{M}$ = 
1.9$\times$10$^{-8}$ $M_\odot$,][]{PHM04} is actually accreted by 
the WD companion.   
In such a case, the modulation observed in the X-ray light curve 
could be associated with a hot spot onto the surface of the WD 
and the period would correspond to its rotational period.  
This scenario, however, is unable to provide an explanation for the 
presence of the O~{\sc vi} $\lambda\lambda$1037,1037 P-Cygni profiles 
in the spectrum of the CSPN, which still demands the production of 
X-rays in its stellar wind \citep{HB11}.

The accretion of stellar wind from the CSPN onto the WD can 
certainly be claimed responsible for the collimation and 
launch of the fast collimated outflows of NGC\,2392.  
However, we consider it very unlikely that these outflows are the source of 
the diffuse X-ray emission confined within the inner shell of NGC\,2392.  
First, the outflow speed of 180 km~s$^{-1}$ is not able 
to shock-heat material up to X-ray-emitting temperatures.  
Second, the spatial distributions of the diffuse X-ray emission and 
collimated outflows do not match;  the collimated outflows extends 
further out from the inner shell \citep{GDetal_2012}, whereas the 
diffuse emission is well confined within the inner shell and it does 
not show any ``hot spot'' indicative of interactions between the 
collimated outflow and the nebular material \citep{Retal_2013}.  
According to recent hot bubble hydrodynamical simulations \citep{TA2016}, the 
stellar wind of the CSPN of NGC\,2392 is only just capable of powering the 
diffuse X-ray emission, a situation that can be favored by the contribution 
of the stellar wind from the WD companion \citep{Metal_2019}.

\section{Summary}

Inspired by the recent discovery of a spectroscopic binary companion 
of the CSPN of NGC\,2392, we have revisited the spectral and temporal 
properties of its hard X-ray emission.  
The X-ray emission is confirmed to be one of the hardest among the 
X-ray emission from CSPNe, with a plasma temperature $\simeq$26 MK 
and an X-ray luminosity close to 10$^{31}$ erg~s$^{-1}$.  
We find evidence that the X-ray emission is modulated, 
with a period of 0.26 days, which is much shorter than 
the 1.9 days orbital period of the companion, but twice 
the 0.123 days rotational period of the CSPN.  
Coronal emission from an unseen companion falls below the levels 
of the observed X-ray emission, whereas the current stellar wind 
is not capable of producing it, either.  
The X-ray luminosity and temperature are consistent with those expected 
from the accretion of material from an unseen main-sequence companion
onto the surface of the CSPN, or from accretion of the CSPN stellar wind
directly onto the surface of a WD companion.  
The former scenario is not supported by the different rotational 
period of the CSPN and time-scale of the modulation of the X-ray 
emission, whereas the latter scenario cannot provide an explanation 
for the O~{\sc vi} UV P-Cygni profiles present in the stellar spectrum,  
which still require an in-situ production of X-rays at the CSPN, maybe 
through shocks in the stellar wind.

The tantalizing variability of the X-ray emission from the CSPN 
of NGC\,2392 provides another important piece of information to 
understand its nature and possible implications on the currently 
active collimation process of its fast outflows.  
A better assessment of this variability and its time-scale 
is necessary for a critical comparison with the orbital 
period of the WD companion and the rotational period of 
the CSPN.  
Upcoming high-sensitivity X-ray missions such as \emph{Athena} 
\citep{Nandra2013} would be able to assess and characterize the 
X-ray variability of the CSPN of NGC\,2392 and that of other 
hard X-ray CSPNe because their emission can be spectrally 
distinguished from that of their soft diffuse emissions.

\acknowledgments

MAG acknowledges financialsupport by grants AYA~2014-57280-P and
PGC2018-102184-B-I00, co-funded with FEDER funds, and from the State
Agency for Research of the Spanish MCIU through the "Center of
Excellence Severo Ochoa" award for the Instituto de Astrof\'\i sica
de Andaluc\'\i a (SEV-2017-0709).  
JAT and MAG are funded by UNAM DGAPA PAPIIT project IA100318.  
YHC is supported by Taiwanese Ministry of Science and Technology
grant MOST 108-2811-M-001-587.
We appreciate Dr.\ F.F.\ Bauer for his help to build the Lomb-Scargle 
periodogram and derive the false alarm probability of the light curve 
of the central star of the Eskimo, and an anonymous referee for useful
comments.  
This research has made use of the SIMBAD database, operated at CDS, 
Strasbourg, France.  
Based on observations made with the Italian Telescopio Nazionale Galileo
(TNG) operated on the island of La Palma by the Fundaci\'on Galileo Galilei 
of the INAF (Istituto Nazionale di Astrofisica) at the Spanish Observatorio 
del Roque de los Muchachos of the Instituto de Astrof\'\i sica de Canarias.



\facility{CXO (ACIS), XMM-Newton (EPIC), IUE, FUSE, ORM TNG (NICS), 
Spitzer (IRAC, IRS)} 

\end{document}